\documentclass[
preprint,
tightenlines
,showpacs
,showkeys
,nofootinbib
,aps
,amsfonts
,amssymb
,pdftex  
]{revtex4-1}

\usepackage{graphicx}  
\usepackage{amsmath, amssymb, bm}   
\usepackage[dvips]{color}
\usepackage{pslatex}  
\usepackage{psfrag,epsfig}
\usepackage{natbib}
\usepackage{hyperref}

\def\){\right)} 
\def\({\left(} 
\def\]{\right]} 
\def\[{\left[}

\def\nLi{$^7\mathrm{Li}(n, \gamma)^8\mathrm{Li}$}
\def\pBe{$^7\mathrm{Be}(p, \gamma)^8\mathrm{B}$}

\def\P { (\frac{\stackrel{\rightarrow}{\nabla}}{M_C}-\frac{\stackrel{\leftarrow}\nabla}{M_N})}

\begin{document}
\preprint{INT-PUB-10-068}

\title{
Radiative Neutron Capture on Lithium-7}

\author{%
Gautam Rupak}
\email{grupak@u.washington.edu}
\affiliation{Department of Physics $\&$ Astronomy and 
High Performance Computing Collaboratory,
 Mississippi State
University, Mississippi State, MS 39762, U.S.A.}

\author{%
Renato Higa}
\email{R.Higa@rug.nl}
\affiliation{Kernfysisch Versneller Instituut, Theory Group, University 
of Groningen, 9747AA Groningen, The Netherlands}

\begin{abstract}
The radiative neutron capture on lithium-7 is calculated model independently 
using a low energy halo  effective field theory.  The cross section 
is expressed in terms of scattering parameters directly related to the 
$S$-matrix element. The cross section depends on the poorly known $p$-wave 
effective range parameter $r_1$. This constitutes the leading order 
uncertainty in traditional model calculations. It is explicitly 
demonstrated by comparing with potential model calculations. 
A single parameter fit describes the low energy data extremely well and
 yields $r_1\approx -1.47$ fm$^{-1}$.

\end{abstract}

 

\maketitle

\section{Introduction\label{sec_intro}}
Low energy nuclear reactions play a crucial role in Big Bang Nucleosynthesis 
(BBN), stellar burning and element synthesis at supernova 
sites~\cite{Burles:1999zt,Rolfs:1988,Barwick:2004ep}. Besides placing 
constraints on our understanding of element formation, these low energy 
reactions play an important role in testing astrophysical models and physics 
beyond the Standard Model of particle physics. Often the key nuclear 
reactions occur at energies that are not directly accessible in terrestrial 
laboratories. Radiative proton capture on beryllium \pBe~ is one 
of them ---it is important for boron-8 production in the sun, whose 
weak decay results in the high energy neutrinos that are detected at 
terrestrial laboratories looking for physics beyond the Standard Model. 
The relevant solar energy, the Gamow peak, for this reaction is around 
$20$ keV~\cite{Adelberger:1998qm}.  This necessitates extrapolation to solar 
energies of known experimental capture cross sections from above 
around $100$ keV. Theoretical input becomes necessary for this extrapolation. 
Effective field theory (EFT) is an ideal formalism for this as it provides a 
model-independent calculation with reliable error estimates.

 In an EFT, one identifies 
the relevant low energy degrees of freedom and constructs the most general 
interactions allowed by symmetry without modeling the short distance physics. 
The interactions are organized in a low momentum expansion. At a given order 
in the expansion, a finite number of interactions has to be considered 
and an {\em a priori} estimate of the theoretical error can be made. 
Establishing theoretical errors is crucial due to astrophysical 
demands~\cite{Burles:1999zt,Rolfs:1988,Adelberger:1998qm}. A systematic 
expansion of interactions is important because many processes involve external 
currents, and any prescription used in phenomenological models involve some 
uncertainty. 
As an example, the cross section for $n(p,\gamma)d$  at BBN energies was 
calculated within EFT to an accuracy of about $1\%$~\cite{Rupak:1999rk}.
Systematic treatment of two-body currents was necessary to achieve this level 
of precision, and it addressed a critical need~\cite{Burles:1999zt} for 
nuclear theory input in astrophysics.

While 
applications of EFT to systems with $A\lesssim 4$ 
nucleons is well developed, for
$A\gtrsim 5$ 
it is still in its 
infancy. However, some loosely bound systems, 
like halo nuclei open new possibilities. The small separation energy of the valence nucleons in halo nuclei provides a small expansion parameter for constructing a halo EFT~\cite{Bertulani:2002sz,*Bedaque:2003wa}.
  The $^8$B nucleus with a proton weakly bound to the $^7$Be core by $0.1375$ MeV is a halo system. 
  Current extrapolation of the \pBe~ cross section to solar energies introduce 
errors in the $5-20\%$ 
range~\cite{Adelberger:1998qm,PhysRevC.68.045802,PhysRevC.70.065802}. 
A model-independent EFT calculation  would be very 
useful to estimate the errors in the extrapolation. 
In addition, this would be an 
important step in developing EFT techniques for weakly-bound nuclei as 
has been accomplished in the few nucleon systems. 
Experiments such as those planned at the future FRIB~\cite{FRIB} would 
explore exotic nuclei near the drip lines where halo systems abound. 
Structures and reactions with halo EFT can serve as benchmark 
for phenomenological models of nuclei near the drip lines.
  
 In this paper we consider the  low energy 
reaction  \nLi, which is a isospin 
mirror to \pBe. The $n$-$^7{\rm Li}$ system 
allows formulating the EFT  for the nuclear interactions 
without the added complication of the Coulomb force. 
Traditionally \nLi~ has been calculated in a single-particle 
approximation as a $^7$Li core plus a valence neutron interacting via a  
Woods-Saxon potential~\cite{Tombrello:1965}. 
This approximation breaks down at higher energies when the internal structure of the $^7$Li core is probed, for example, near the threshold for $^7$Li$(\gamma, ^3$He$)\alpha$ which is  about $0.5$ MeV above 
the binding energy $B\approx 2.03$ MeV of the $^8$Li core. 
We treat  the  $^7$Li nucleus as point-like since we work at very low energies. 
Once the nuclear piece is calculated in EFT for the 
$n$-$^{7}{\rm Li}$ 
system the Coulomb interaction in $p$-$^{7}{\rm Be}$ can be incorporated 
systematically as have been done for proton fusion in 
EFT~\cite{Kong:1999tw,*Kong:1999sf,*Kong:1999mp,*Kong:1998sx,*Kong:2000px}. 
The reaction \nLi, besides being a check on the mirror 
\pBe~reaction, is important in inhomogeneous BBN. It impacts the production of carbon-oxygen-nitrogen in the
early universe, and constrains astrophysical models~\cite{kawano:1991ApJ372}. 
We calculate the \nLi~ reaction analytically and express the result in 
terms of parameters directly related to observables, thus quantifying 
the dominant theoretical uncertainty in the single particle 
approximation.

\section{Interaction\label{sec_interaction}}
The relevant low energy nuclear degrees of freedom, here, are the point-like neutron, $^7$Li and $^8$Li with spin-parity 
$\frac{1}{2}^+$, $\frac{3}{2}^-$ and $2^+$ respectively.  
At low energies the relevant  
partial waves in the incoming neutron-lithium state are $s$-waves: 
$^3S_1$, $^5S_2$ in the spectroscopic notation $^{2S+1}L_J$. 
The ground state is a $2^+$ state that is primarily 
the symmetric combination of the possible $p$-wave states $^3P_2$ and 
$^5P_2$~\cite{Trache:2003}. Conservation of parity implies that the reaction \nLi~ proceeds through the electric dipole transition E1 at lowest order.

It is known that the non-relativistic amplitude in the $l$-th partial wave has the 
general form
 \begin{align}\label{eq:ERE}
i \mathcal A_l(p) =\frac{2\pi}{\mu}
\frac{i p^{2 l}}{p^{2 l+1} \cot\delta_l-i p^{2l+1}}, 
 \end{align} 
where $\mu$ is the reduced mass of the $n$-$^7{\rm Li}$  system 
with masses $M_N$ and $M_C$, respectively, and $\delta_l$ is the partial wave 
phase shift. 
The term $p^{2l+1} \cot\delta_l = -1/a_l+r_l p^2/2 +\cdots$ has an analytic 
effective range expansion (ERE) for short range interactions.
Scattering and bound 
state information in EFT are incorporated by matching the EFT couplings 
 to the ERE parameters  in the low energy 
expansion.

For the initial $s$-wave states, at sufficiently low momentum 
 $i\mathcal A_0\approx -i \frac{2\pi}{\mu} a_0$, 
and  one keeps only the first term in the ERE, corresponding 
to a single perturbative insertion of the leading EFT interaction. 
However, to describe  shallow  bound or 
virtual states that correspond to large scattering length $a_0\gg r_0$ one has to expand around the $1/a_0$ pole term and write 
\begin{align}\label{eq:ERE2}
i \mathcal A_0 \approx i \frac{2\pi}{\mu}\frac{1}{-\frac{1}{a_0}-i p}
[ 1- \frac{r_0 p^2}{2(-\frac{1}{a_0}-ip)}+\cdots]. 
\end{align}
This requires a non-perturbative resummation of a single interaction in EFT at leading order. Such a resummation extends the validity of the EFT to include the shallow state at momenta $p\sim 1/a_0$.  
In the $n$-$^7$Li system, the scattering length is 
$a^{(2)}_0= -3.63\pm0.05$ fm ($a^{(1)}_0=0.87\pm0.07$ fm) in the $^5S_2$ 
($^3S_1$) spin-channel~\cite{Koester:1983,*Angulo:2003}. 
This corresponds to neutron momentum around $54$ ($227$) MeV, or center of 
mass (CM) energy $2$ ($31$) MeV. 
We are interested at the extremely low solar energies with momenta 
$p\ll 54$ MeV. Thus in  EFT a single perturbative interaction in each 
of the $^5S_2$ and $^3S_1$ channels is required.  The interaction can be 
resummed in the $^5S_2$ channel if one wants to compare with data at CM 
energies $\sim 1$ MeV.

The leading order interactions for $s$-wave contain no derivatives. 
The spin-$\frac{1}{2}$ neutron and spin-$\frac{3}{2}$  $^7$Li nucleus can be 
combined into the $^3S_1$  and $^5S_2$ states using the 
Clebsch-Gordan coefficient matrices $F_i$, $Q_{i j}$ as $N^T F_i C$ and 
$N^T Q_{i j} C$ respectively. 
The vector index in $F_i$ relates to the three magnetic quantum numbers in 
the spin $S=1$ channel. The symmetric, traceless matrices $Q_{ij}$ relate to 
the five  magnetic quantum numbers in the spin $S=2$ channel.  We write the 
$s$-wave interaction Lagrangian as
\begin{align}\label{eq:Ls}
\mathcal L^{(s)}=g^{(1)}(N F_i
C)^\dagger(N F_i C)+g^{(2)} (N Q_{i j} C)^\dagger (N Q_{i j} C)+\cdots,
\end{align} 
where a single momentum-independent interaction in each of the $^3S_1$ and $^5S_2$ 
channels  was kept. 
The ``$\cdots$" represents higher derivative terms that are suppressed at 
low energy. The $2\times4$ Clebsch-Gordan matrices are given as
\begin{align}
F_i=&-\frac{i\sqrt{3}}{2}\sigma_2 S_i, 
&{}
&Q_{i j} = -\frac{i}{\sqrt{8}}\sigma_2[\sigma_i S_i+\sigma_j S_i], \\
S_1=&\frac{1}{\sqrt{6}}\(\begin{array}{cccc}
-\sqrt{3} & 0 & 1 & 0\\
0&-1&0&\sqrt{3}
\end{array}\) , \,\,
&{} 
&S_2 = -\frac{i}{\sqrt{6}} \(\begin{array}{cccc}
\sqrt{3} & 0 & 1 & 0\\
0&1&0&\sqrt{3}
\end{array}\) , \,\,
S_3 = \frac{2}{\sqrt{6}} \(\begin{array}{cccc}
0 & 1 & 0 & 0\\
0&0&1&0
\end{array}\) . \nonumber
\end{align}

\begin{figure}[tbh]
\begin{center}
\includegraphics[width=0.48\textwidth,clip=true]{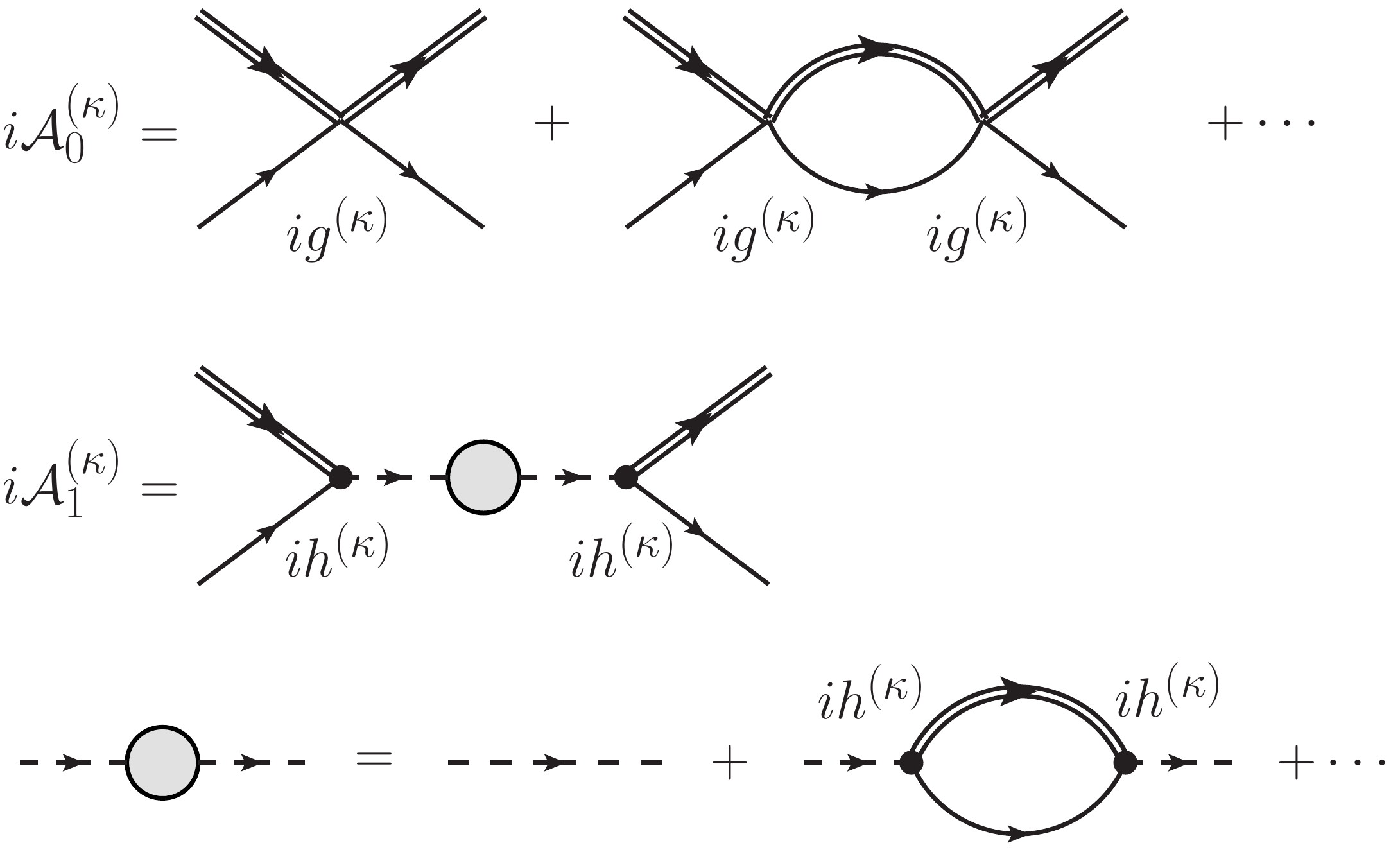} 
\end{center}
\caption{\protect $\mathcal A_0^{(\kappa)}$ is the  $^3S_1$,  $^5S_2$  scattering amplitude. 
 $\mathcal A_1^{(\kappa)}$ is the $^3P_2$, $^5P_2$ scattering amplitude. Double line is the $^7$Li propagator, single line the neutron 
propagator, dashed line the bare dimer propagator.}
\label{fig:scattering}
\end{figure}

The interaction in Eq.~(\ref{eq:Ls}) produces a $s$-wave amplitude shown in 
Fig.~\ref{fig:scattering}. It becomes a geometric series that is 
summed to give 
\begin{align}\label{eq:EFT}
&i\mathcal A^{(\kappa)}_{EFT}(p)=\frac{i g^{(\kappa)}}{1-i g^{(\kappa)} L(p)},
\\
&L(p)= -i 2\mu\(\frac{\lambda}{2}\)^{4-D}\int \frac{d^{D-1}\bm{q}}{(2\pi)^{D-1}}\frac{1}{q^2- p^2-i 0^+}= -\frac{i\mu}{2\pi}(\lambda+ip), \nonumber 
\end{align}
where 
$g^{(\kappa)}$ corresponds to $g^{(1)}$, $g^{(2)}$ in the respective spin channels and 
$\lambda$ is the renormalization scale. The loop integral 
$L(p)$ is evaluated in
the power divergence subtraction scheme~\cite{Kaplan:1998tg} where 
divergences in both $D=4$ and lower space-time dimensions are subtracted. 
Matching Eqs.~(\ref{eq:ERE2}) and (\ref{eq:EFT}) fixes the EFT couplings as 
$g^{(\kappa)}(\lambda)= (2\pi)/[{\mu}(\lambda-1/a^{(\kappa)}_0)]$. 
Introduction of the renormalization scale $\lambda$  allows for a 
systematic expansion of the different terms even though the final amplitude 
is independent of $\lambda$~\cite{Bedaque:1997qi,*Chen:1999tn}. 
In Ref.~\cite{Typel:2004us}, initial state interactions using ERE was also considered.

The $^8$Li nucleus in the final state of the reaction \nLi~ is in 
$p$-wave. We will treat it as a shallow bound state similar to its 
isospin mirror $^8$B nucleus. 
The EFT for a shallow $p$-wave  bound state was formulated in 
Ref.~\cite{Bertulani:2002sz,*Bedaque:2003wa} 
where it was shown that, 
unlike $s$-wave, 
 it requires not one but two
non-perturbative EFT interactions.  The consistent renormalization of loops 
is easily accomplished in the dimer formalism where four-fermion 
interactions are rewritten in terms of a spin-2 dimer and neutron-core 
interactions. The interactions in the $^3P_2$ and $^5P_2$ state can be 
constructed by combining the  matrices $F_i$, $Q_{i j}$ and the Galilean 
invariant velocity difference vector $(\bm{v}_C-\bm{v}_N)_k$ into a 
$p$-wave state with total $J=2$.  We write the 
$p$-wave interaction Lagrangian as 
\begin{align}
\mathcal L^{(p)}=&
\phi_{i j}^\dagger\[\Delta^{(1)} +\(i\partial_0+\frac{\nabla^2}{2 M}\) \]\phi_{i j} 
+ h^{(1)}\sqrt{3} \[\phi_{i j}^\dagger N F_x \P_y C +h.c.\]  R_{ijxy}\\
&+\pi_{i j}^\dagger\[\Delta^{(2)} +\(i\partial_0+\frac{\nabla^2}{2 M}\) \] \pi_{i j} 
+\frac{h^{(2)}}{\sqrt{2}} \[\pi_{i j}^\dagger N Q_{x y} \P_z C +h.c.\] 
T_{xyz ij},\nonumber
\end{align}
where $\phi_{i j}$ ($\pi_{i j}$) is the dimer in the $^3P_2$ ($^5P_2$) 
channel, and
\begin{align}
R_{ijxy}&=\frac{1}{2}[\delta_{i x}\delta_{j y}+\delta_{i y}\delta_{j x} -\frac{2}{3}\delta_{i j}\delta_{x y}], \ \ \ \
T_{xyz i j}=\frac{1}{2}\[\epsilon_{x z i}\delta_{y j}+\epsilon_{x z j}\delta_{y i}
+\epsilon_{y z i}\delta_{x j}+\epsilon_{y z j}\delta_{x i} \] . 
\end{align}
The interactions in $\mathcal L^{(p)}$ are equivalent to the ones
with only neutron-core short range interactions without a dimer field. 
In terms of Feynman diagrams, the  
four-fermion neutron-core 
interaction is replaced in the dimer formulation by a dimer exchange, 
Fig.~\ref{fig:scattering}. 
The non-perturbative iteration of the leading operators is accomplished 
by ``dressing" the dimer propagator with nucleon-core loops. 
For a given spin-channel $\kappa=1$ ($^3P_2$), $2$ ($^5P_2$) the dressed 
dimer propagator, which is proportional to the elastic amplitude, reads
\begin{align}
i D^{(\kappa)}(p_0,\bm{p})R_{i j m n}=&  \frac{i R_{i j m n} }
{\Delta^{(\kappa)} - \frac{1}{2\mu}\zeta^2 +\frac{2h^{(\kappa)\,2}}{\mu}
f (p_0,\bm{p})},\\
f(p_0,\bm{p})=&
\frac{1}{4\pi}\left(\zeta^3-\frac{3}{2}\zeta^2\lambda
+\frac{\pi}{2} \lambda^3 \right),
\nonumber
\end{align}
where $\zeta=\sqrt{-2\mu p_0 +\mu p^2/M -i 0^+}$, $M=M_N+M_C$. 
Matching the EFT amplitudes to the $p$-wave  ERE expansion 
determines the  coupling pair 
($\Delta^{(\kappa)}$, $h^{(\kappa)}$). 
Again, only the first two ERE parameters are kept in the low energy 
expansion since EFT requires two operators at leading order.

\section{Radiative capture\label{sec_capture}}
\begin{figure}[tbh]
\begin{center}
\includegraphics[width=0.47\textwidth,clip=true]{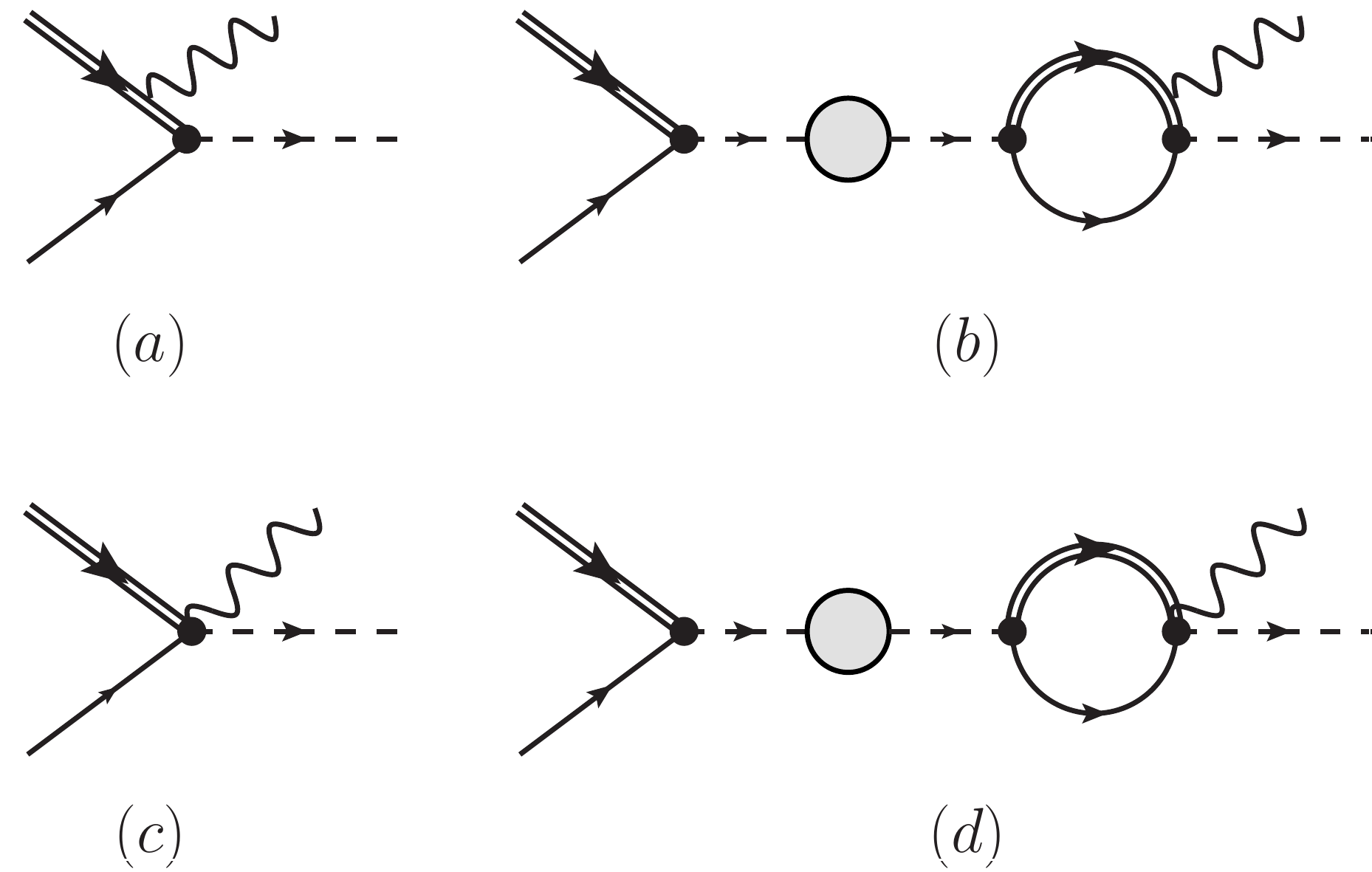} 
\end{center}
\caption{\protect  Capture reactions \nLi.  Wavy lines represent photons. }
\label{fig:capture}
\end{figure}

The leading order capture cross section can be calculated via minimally 
coupling the photon by gauging the $^7$Li core momentum 
$\bm{p}\rightarrow  \bm{p} + Z_{C} e \bm{A}$, where $Z_C=3$ is the 
$^7$Li core charge. 
The E1 contribution to the cross section comes from the diagrams in 
Fig.~\ref{fig:capture}. The CM kinematics are defined as: $\bm{p}$ the 
core momentum, $\bm{k}$ the photon momentum and 
$\hat{\bm k}\cdot\hat{\bm p}=\cos\theta$. 
Formally 
we take $p\sim\gamma$ as the small scale  where 
$\gamma=\sqrt{2\mu B}\approx 57.8$ MeV is the $^8$Li binding momentum. 
Then at leading order the Mandelstam variable $s\approx (M_N+M_C)^2= M^2$ 
and $|\bm{k}|=k_0\approx(p^2+\gamma^2)/(2\mu)$. 
We get for the CM differential cross section 
\begin{align}
\frac{d\sigma}{d\phi d \cos\theta}=\frac{1}{64\pi^2 s}\frac{|\bm{k}|}{|\bm{p}|}|\mathcal M|^2
\approx \frac{1}{64\pi^2  M^2}\frac{p^2+\gamma^2}{2\mu p}|\mathcal M|^2. 
\end{align}

The capture from the initial state $^5S_2$ to the $^5P_2$ final state 
(spin channel 2) dominates due to the larger initial state scattering 
length $a_0^{(2)}>a_0^{(1)}$.   
 The divergence in diagram $(b)$ is canceled by $(d)$.
  Summing over all polarizations and spins we get
\begin{align}\label{eq:5p2capture}
|\mathcal M^{(^5P_2)}|^2=&
\frac{5|Z_{\pi}|\left[8\pi Z_C h^{(2)}\right]^2\alpha M M_n}{\pi M_c}
\[ (1+ X)(1+X^\ast) -\frac{p^2\sin^2\theta}{p^2+\gamma^2}\(\frac{2\gamma^2}{p^2+\gamma^2}+ X +X^\ast\) \],\\
X=&\frac{i}{-1/a^{(2)}_0-i p}
(p-i\frac{2}{3}\frac{\gamma^3-ip^3}{p^2+\gamma^2})\nonumber,  
\end{align}
with the dimer polarization sum 
$\sum \varepsilon_{i j}\varepsilon_{xy}^\ast=R_{ijxy}$
~\cite{Choi:1992,*Fleming:1999ee} and  
the wave function renormalization 
$h^{(2)\,2}|Z_\pi|=2\pi/|3\gamma+r^{(2)}_1|$,
where $r^{(2)}_1$ is the effective range in the $^5P_2$ scattering amplitude. 
$Z_\pi$ is defined as the residue at the pole in the dressed dimer propagator 
$D_\pi(p_0,\bm{p})$~\cite{Brown:book}. 
The capture from the $^3S_1$ state to the $^3P_2$ state has the same exact 
expression as Eq.~(\ref{eq:5p2capture}) except that $a^{(2)}_0$ and $r^{(2)}_1$ are replaced by the corresponding 
parameters in the 
spin channel 1.
The differential cross section averaged over initial spin states is 
\begin{align}\label{eq:totalsigma}
\frac{d\sigma}{d \cos\theta}=\frac{1}{32\pi  M^2}
\frac{p^2+\gamma^2}{2\mu p}\frac{1}{8}
\frac{|\mathcal M^{(^5P_2)}|^2+|\mathcal M^{(^3P_2)}|^2}{2}, 
\end{align}
taking the $^8$Li nucleus to be a symmetric combination 
$(|^3P_2\rangle+|^5P_2\rangle)/\sqrt{2}$ of final states.
The total cross section $\sigma(p)$ is calculated with a straightforward 
integration over the angle $\theta$.  

The parameters in $\sigma(p)$ can be determined from elastic $n$-$^7$Li 
scattering data and $^8$Li binding energy. However, the $p$-wave effective 
range $r^{(\kappa)}_1$ is not known accurately. This is the main theoretical 
uncertainty at this order. 
Changing the effective range $r^{(\kappa)}_1$ modifies the 
wave function renormalization factor and moves the cross section up or down 
by a constant factor. In traditional potential model calculations, the 
parameters are determined by reproducing the $^8$Li binding energy. 
However, this does not constrain the effective range and other parameters 
of the ERE. For example, in a Woods-Saxon potential 
$V(r) = - v_0[1+\exp(\frac{r-R_c}{a_c})]^{-1}$ different choices for the 
depth $v_0$, range $R_c$, diffusiveness $a_c$ can be made to reproduce the 
known $^8$Li binding energy. This however produces different effective 
ranges, and constitutes an irreducible source of error in the theoretical
calculations. 

Comparing the contributions to the capture cross section from the two 
spin channels analytically, we get   
\begin{align}
\frac{\sigma^{(^5P_2)}}{\sigma^{(^5P_2)}+\sigma^{(^3P_2)}}\Big|_{p=0}=\frac{(3-2a_0^{(2)}\gamma)^2}{
(3-2a_0^{(2)}\gamma)^2+(3-2a_0^{(1)}\gamma)^2}\approx 0.81,
\end{align}
using the same effective range $r_1$ in both 
spin channels. 
This ratio is close to the experimentally observed ratio~\cite{Barker:1995a}. 
From Eqs.~(\ref{eq:5p2capture}),~(\ref{eq:totalsigma}) one can see that the total cross section at low energy is not 
independently sensitive to $r_1^{(2)}$ and $r_1^{(1)}$. 
This is confirmed by our fit to data.

\begin{figure}[tbh]
\begin{center}
\includegraphics[width=0.48\textwidth,clip=true]{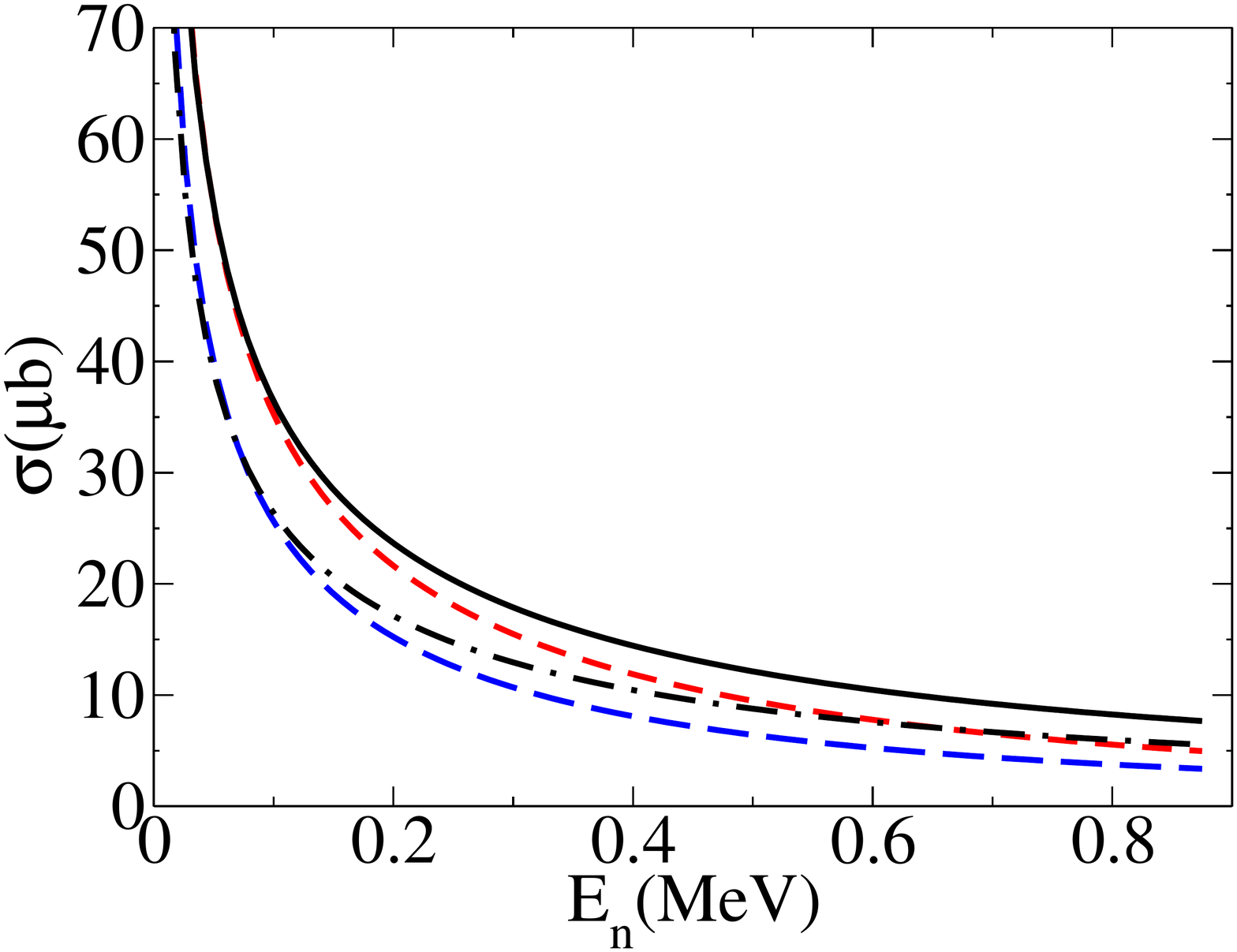} 
\includegraphics[width=0.48\textwidth,clip=true]{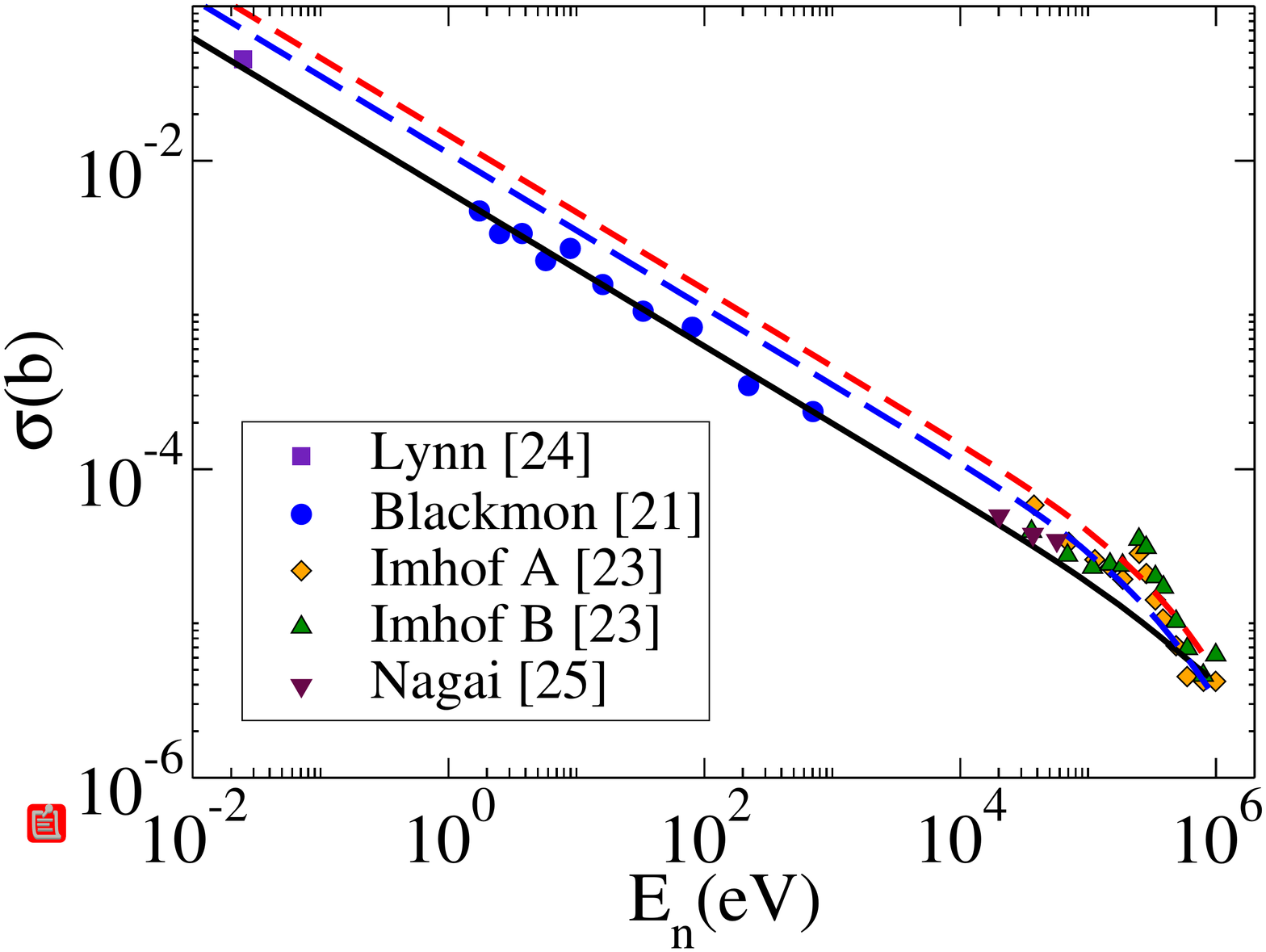} 
\end{center}
\caption{\protect Potential model curves: (blue) long-dashed curve
from Davids-Typel~\cite{PhysRevC.68.045802}, (red) dashed curve from Tombrello~\cite{Tombrello:1965}.  
Left panel:  (black) solid curve EFT with 
$r_1=-0.46$ fm$^{-1}$, (black) dot-dashed curve  EFT with $r_1=-0.3$ 
fm$^{-1}$. 
Right panel: (black) solid curve EFT with  $r_1$ fitted to 
data. }
\label{fig:theory}
\end{figure}

In Fig.~\ref{fig:theory}, we compare potential model calculations using 
Tombrello's~\cite{Tombrello:1965}, and 
Davids-Typel's~\cite{PhysRevC.68.045802} parameters to EFT curves. 
At low energy the potential model results can be reproduced in EFT with 
a small variation in the effective range 
$-0.46$ fm$^{-1}\leq r_1\leq -0.3$ fm$^{-1}$. 
At higher energies they differ since potential models include ERE parameters 
beyond the scattering length and effective range. 
A fit to data from Ref.~\cite{Blackmon:1996} in the energy range 
$E_n\sim 2-700$ eV gives an effective range $r_1=-1.83$ fm$^{-1}$ with 
only the spin channel $2$ contribution and $r_1=-1.47$ fm$^{-1}$ with both 
spin channels $1$ and $2$.  
Both the $r_1$ 
values are compatible with the Wigner bound~\cite{Hammer:2009zh,*Hammer:2010fw} 
which, for a nucleon-core interaction shorter than 3 fm restricts $r_1$ 
to be smaller than around $-1$ fm$^{-1}$. 
Following Ref.~\cite{Blackmon:1996}, their data and the theory curves in  the right panel in Fig.~\ref{fig:theory} were divided by the known experimental branching ratio $0.89$ to the ground state and compared to a few other 
available data~\cite{Imhof:1959,Lynn:1991,Nagai:2005}.  The $r_1$ was fitted to the 
unscaled data for transition to the ground state as appropriate. 
It is clear that the theory 
error in the low energy extrapolation comes from the uncertainty in the 
effective range at leading order.

\section{Conclusions\label{sec_conclusion}}
We considered radiative capture reactions for halo nuclei. 
The low energy \nLi~ cross section was calculated at leading order using EFT. 
In the single particle approximation, the cross section was derived in terms 
of scattering parameters that are directly related to $S$-matrix elements. 
Using a model-independent formalism we demonstrated and quantified the 
theoretical uncertainty associated with phenomenological potentials in the 
single particle approximation. 
The leading order result depends on the $p$-wave effective range parameter 
$r_1$ that is poorly known. 
Without detailed knowledge about this parameter, model calculations deviate 
from data at low energy. We extract the effective range $r_1$ by fitting our 
analytic form to data. 
 
At higher order in the EFT expansion, the cross section would get 
corrections from two sources: higher order initial and final state 
interactions, and two-body currents. 
The initial and final state interactions can be related to the ERE. 
At the very low energy, it is the final state interactions, 
which modify the wave function renormalization constants, that are 
important. 
At next-to-next-to-leading order the shape parameter associated with $p$-wave 
interaction contribute~\cite{Bertulani:2002sz,*Bedaque:2003wa, HigaRupak}. In addition, at higher order two-body currents 
such as $E_i (N F_j C)^\dagger [NF_x( 
\stackrel{\rightarrow}{\nabla}/M_C - \stackrel{\leftarrow}{\nabla}/M_N)_y C]  R_{ijxy}$, where $E_i$ is the 
electric field, contribute. These operators are not constrained by elastic 
scattering. 
A higher order EFT calculation would reduce theoretical errors though at 
the expense of additional parameters. This is not necessarily a drawback 
as what we gain is a model-independent  understanding of the sources of 
higher order contributions, and a more detailed knowledge about the kind 
of experimental input that is required to better constrain the low energy theory. 

Coulomb interactions in $p+{}^7$Be scattering and \pBe~reaction is being 
considered where the current formulation plays a crucial 
role~\cite{HigaRupak}. The power counting of 
electromagnetic currents beyond leading order is being considered as well.

\acknowledgments
The authors thank P. Bedaque,   C. Bertulani,  B. Davids, C. Johnson, 
A. Mukhamedzhanov, S. Typel  for valuable discussions. 
Authors are extremely grateful  to S. Typel  for providing the potential 
model numbers. 
Authors thank ECT* and INT, and R.H. thanks MSU for hospitality  
where part of this research was performed. 
The work of G.R. is partially supported by the HPCC center at MSU and the 
U.S. NSF grant PHY-0969378.
The work of R.H. was partially supported by the Dutch Stichting voor 
FundamenteelOnderzoek der Materie under programme 104 and by the BMBF 
under contract number 06BN411. 


%

\end{document}